\documentclass[twocolumn,prb,amsmath,amssymb]{revtex4}
\usepackage{graphicx}% Include figure files
\usepackage{booktabs}
\usepackage{multirow}
\usepackage{amssymb}
\usepackage[version=3]{mhchem} % Formula subscripts using \ce{}
\usepackage{graphicx}
\usepackage{amsmath}
\usepackage{indentfirst}
\usepackage{pstricks}
\usepackage{verbatim}
\usepackage{bm}
\usepackage{color}   % colors

\def\be{\begin{equation}}
\def\ee{\end{equation}}

\begin{document}
\title
{Tm-doped fiber laser mode-locked by graphene-polymer composite}
\author{M. Zhang$^{1}$, E.J.R. Kelleher$^{1}$, F. Torrisi$^2$, Z. Sun$^2$, T. Hasan$^2$,\\ D.Popa$^2$, F. Wang$^2$, A.C. Ferrari$^2$, S.V. Popov$^1$ and J.R. Taylor$^{1}$}

\affiliation{$^1$Femtosecond Optics Group, Department of Physics, \\ Imperial College London, Prince Consort Road, London SW7 2BW, UK
\\
$^2$Department of Engineering, University of Cambridge, Cambridge, CB3 0FA, UK}

\begin{abstract}
We demonstrate mode-locking of a thulium-doped fiber laser operating at 1.94$\mu$m, using a graphene-based saturable absorber. The laser outputs 3.6ps pulses, with$\sim$0.4nJ energy and an amplitude fluctuation$\sim0.5\%$, at 6.46MHz. This is a simple, low-cost, stable and convenient laser oscillator for applications where eye-safe and low-photon-energy light sources are required, such as sensing and biomedical diagnostics.
\end{abstract}
\maketitle

Ultrafast lasers based on thulium (Tm) doped fibers, operating in the 1.7-2.1$\mu$m range, are important to address demands for mid-IR sources necessary for a variety of applications, ranging from molecular spectroscopy\cite{Nelson:95} and biomedical diagnostics\cite{Sharp:96}, to medicine\cite{Kivisto:07} and remote sensing\cite{Solodyankin:08,Wang_ol_2009}. The 2$\mu$m region is important because several gas molecules (e.g. CO$_2$) have characteristic absorption lines there\cite{sensor10}. Since liquid water (main constituent of human tissue) absorbs more strongly at$\sim$2$\mu$m ($\sim$100/cm)\cite{LP-walsh-2009} than at$\sim$1.5$\mu$m ($\sim$10/cm)\cite{LP-walsh-2009} and$\sim$1$\mu$m ($\sim$1/cm)\cite{LP-walsh-2009}, laser sources emitting at$\sim$2$\mu$m are promising for medical diagnostic and laser surgery\cite{LP-walsh-2009}. In addition, Light Detection And Ranging\cite{Ebrahim_book} measurements and optical free-space telecommunications\cite{Ebrahim_book} can be performed within the 2-2.5$\mu$m atmospheric transparency window\cite{AO-Gebhardt-1976}. Furthermore, fiber lasers offer advantages compared to solid-state lasers, such as compact geometry, efficient heat dissipation and alignment-free operation\cite{Fermann_jstqe_09,Okhotnikov_njp_2004}.

2$\mu$m fiber lasers have been mainly mode-locked using nonlinear polarization evolution (NPE)\cite{Nelson:95} and semiconductor saturable absorber mirrors (SESAMs)\cite{Kivisto:07}. However, these have disadvantages: NPE suffers from bulky construction and environmental sensitivity\cite{Nelson:95}, SESAMs have complex fabrication and packaging, as well as limited bandwidth\cite{Keller_nature_03}. Nanotubes and graphene are promising saturable absorbers (SA) due to their favorable properties: ultrafast recovery time\cite{Chen_apl_2002,Breusing_prl_2009,Brida_cleo_2012}, broadband operation\cite{Wang_nn_2008,Sun_tun}, ease of fabrication\cite{Hasan:09,Wang_nn_2008,Sun_tun,Popa_apl_11,Bonaccorso:10} and integration\cite{Hasan:09,Bonaccorso:10} into all-fiber configurations. On one hand, broadband operation can be achieved using a distribution of nanotube diameters\cite{Wang_nn_2008}. On the other hand, this is an intrinsic property of graphene, due to the gapless linear dispersion of Dirac electrons\cite{Sun:10,Sun_tun,Popa_apl_11,Bonaccorso:10}. Nanotubes have mode-locked fiber\cite{Wang_nn_2008,Set_jstq_2004,Kelleher_ol_09,Scardaci_am_2008,Sun_apl_2008,Sun_apl_2009,Sun_nr_2010}, waveguide\cite{Dellavalle_apl_2006,Beecher_apl_2010}, solid-state\cite{Schibli_oe_2005,Schmidt_ol_2008,Obraztsov_or_2010} and semiconductor lasers\cite{Song_ol_2007}, covering from$\sim$0.8 to$\sim$2$\mu$m\cite{Hasan:09,Bonaccorso:10}. Ultrafast pulse generation at 0.8\cite{Baek_ape_2012}, 1\cite{Tan_apl_10}, 1.3\cite{Cho_ol_08}, 1.5$\mu$m\cite{Hasan:09,Bonaccorso:10,Sun:10,Sun_tun,Popa_2010} was demonstrated by exploiting graphene saturable absorbers (GSAs). Refs.\onlinecite{Liu_lpl_11,Ma_ol_2012} reported 2$\mu$m solid-state lasers mode-locked with graphene oxide and chemical vapor deposited (CVD) 1-2 layer graphene. However graphene oxide\cite{stankovich_nat_06,Mattevi_afm_09} is fundamentally different from graphene; it is an insulating material with a mixture of sp$^2$/sp$^3$ regions\cite{stankovich_nat_06,Mattevi_afm_09}, with lots of defects and gap states\cite{Mattevi_afm_09}. Thus it does not offer in principle the wideband tunability of graphene. CVD graphene, on the other hand, is normally grown at very high temperature on Cu\cite{Li_s_2009} or Ni\cite{Kim_n_2009}. Therefore, extra steps are required to transfer graphene to the target substrates for photonic applications. Indeed, graphene can be produced in a variety of ways, ranging from micromechanical cleavage\cite{Bonaccorso:10}, to liquid phase exfoliation(LPE)\cite{Hernandez08}, CVD of hydrocarbons\cite{bae_nn_10,Li_s_2009}, carbon segregation from silicon carbide\cite{berger_dheer_2004} or metal substrates\cite{sutter_natmat_07} and chemical synthesis from polyaromatic hydrocarbons\cite{wu_cr_07}. LPE has the advantage of scalability, room temperature processing, and does not require any growth substrate. This produces dispersions that can be easily embedded into polymers to form composites with novel optoelectronic properties to be integrated into various systems\cite{Bonaccorso:10}.
\begin{figure}
\centerline{\includegraphics[width=90mm]{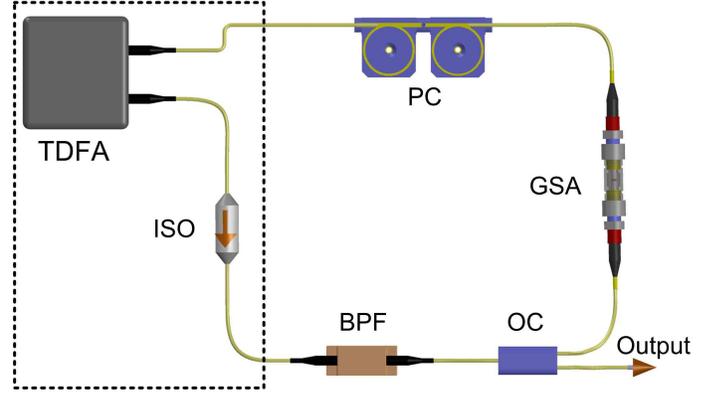}}
\caption{\label{setup} Scheme of our laser cavity. TDFA-Tm-doped fiber amplifier; ISO-isolator; BPF-bandpass filter; OC-output coupler; GSA-graphene-SA; PC-polarization controller}
\end{figure}

Here we demonstrate a fiber laser mode-locked using a graphene-polymer composite. It operates at$\sim$2$\mu$m, with low-noise 3.6ps pulses. Our results show the potential of GSAs for practical fiber lasers in mid-IR.

The laser cavity is schematically shown in Fig.\ref{setup}. It comprises all-fiber integrated components to create an environmentally robust and compact system. A Tm-doped fiber amplifier (TDFA), with integrated optical isolator (ISO), having$\sim$25dB small signal gain at 1.94$\mu$m, and a broad gain bandwidth (full width at half maximum, FWHM$\sim$60nm) is followed by a fiber pigtailed airgap ($\sim$80\% insertion loss) used to include a bandpass filter (BPF) for pulse stabilization, with 80\% maximum transmission and 11nm transmission bandwidth, centered$\sim$1.94$\mu$m. A fused-fiber output coupler (OC) extracts 10\% of the light per pass; a polarization controller (PC) allows adjustment of the intra-cavity polarization.

The graphene-polymer composite is produced as follows: 120mg of graphite (NGS, Naturgraphit) and 90mg sodium deoxycholate (SDC) are sonicated in at room temperature. The unexfoliated particles are allowed to settle for 10 minutes, followed by 60 min centrifugation at$\sim$17000g. The top 70\% of the centrifuged dispersion is then used for the composite fabrication. Drops are also placed on Transmission Electron Microscope (TEM) grids for analysis in a high resolution TEM (HRTEM). Combined HRTEM and normal-incidence/tilted angle electron diffraction measurements show that our dispersion has$\sim$66\% $\leq$ 3-layer flakes ($\sim$26\% single layer,$\sim$22\% bi-layer and$\sim$18\% tri-layer). The remainder have less than 10 layers. The dispersion is also drop-cast on Si/SiO$_{2}$ for Raman measurements with a Renishaw 1000. 5ml of dispersion is then mixed with polyvinyl alcohol (PVA) in water ($\sim$2wt\%) and centrifuged at$\sim$4000g. Evaporation at room temperature gives a$\sim$40$\mu$m film, then used for Raman and absorption measurements.

\begin{figure}
\centerline{\includegraphics[width=90mm]{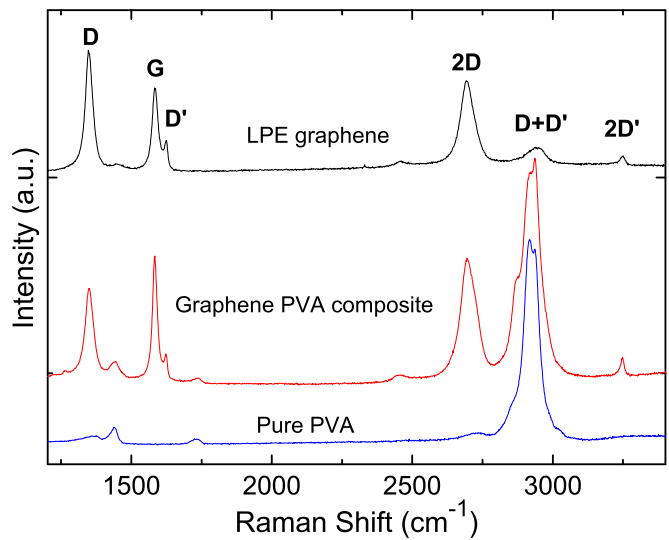}}
\caption{\label{raman} Raman spectra of flake on Si/SiO$_{2}$, polyvinyl alcohol (PVA), graphene-PVA composite}
\end{figure}

Fig.\ref{raman} plots a typical Raman spectrum of a flake deposited on Si/SiO$_{2}$. Besides the G and 2D peaks, this has significant D and D' intensities\cite{Ferrari2000,ferrari_prl_06}. We assign the D and D' peaks to the edges of the submicrometer flakes, rather than a large amount of disorder within the flakes\cite{casiraghi09}. This is further supported by analyzing the G peak dispersion, Disp(G). In disordered carbons the G peak position, Pos(G), increases with decreasing excitation wavelength, from IR to UV\cite{Ferrari2000}. Thus, Disp(G)=$\Delta Pos(G)/\Delta \lambda_{\rm L}$, where $\lambda_L$ is the laser excitation wavelength, increases with disorder\cite{Ferrari2001,Ferrari2000}. FWHM(G) always increases with disorder\cite{Cancado_nl_2011}. Hence, combining the intensity ratio of the D and G peaks, I(D)/I(G), with FWHM(G) and Disp(G) allows us to discriminate between edges, and disorder in the bulk of the samples. In the latter case, a higher I(D)/I(G) would correspond to higher FWHM(G) and Disp(G). By analyzing 30 flakes, we find that the distribution of Disp(G), I(D)/I(G) and FWHM(G) are not correlated, indicating that the D peak is mostly due to edges. Also, Disp(G) is nearly zero for all samples (compared to$\geq$0.1cm$^{-1}$/nm expected for disordered carbons\cite{Ferrari2001}). Although 2D is broader than in pristine graphene, it is still a single Lorentzian. This implies that, even if the flakes are multilayers, they are electronically decoupled and, to a first approximation, behave as a collection of single layers\cite{latil_prb_2007}. Fig.\ref{raman} compares a typical flake with our graphene-PVA composite and pure PVA. We note that the spectrum of the composite (Fig.\ref{raman}) is a superposition of that of the flake and PVA . Thus, PVA does not affect the structure of the embedded flakes.
\begin{figure}
\centerline{\includegraphics[width=90mm]{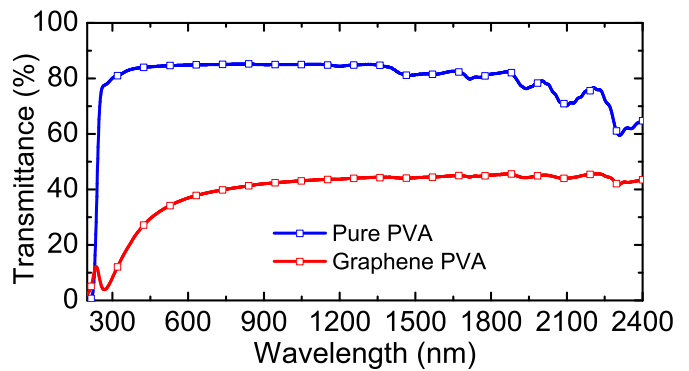}}
\caption{\label{LinTrans} Transmittance of PVA and graphene-composite.}
\end{figure}

Fig.\ref{LinTrans} plots the transmittance of graphene-PVA compared to pure PVA. The UV peak in graphene-PVA is a signature of the van Hove singularity in the graphene density of states\cite{kravetsprb10}. Strong UV absorption is also observed in pure PVA\cite{Haas_jpsp_1963}. By considering the pure PVA absorption, we can estimate that of the graphene component to be$\sim$50\% in the NIR. Given that a monolayer absorbs$\sim$2.3\%\cite{Nair_s_08}, we estimate that an average$\sim$21 layers cross the light path.

The GSA is then prepared by sandwiching 2mm$^2$ of the composite between two fiber connectors, adhered with index matching gel. The integrated device has$\sim$4dB ($\sim$60\%) total insertion loss.
\begin{figure}
\centerline{\includegraphics[width=90mm]{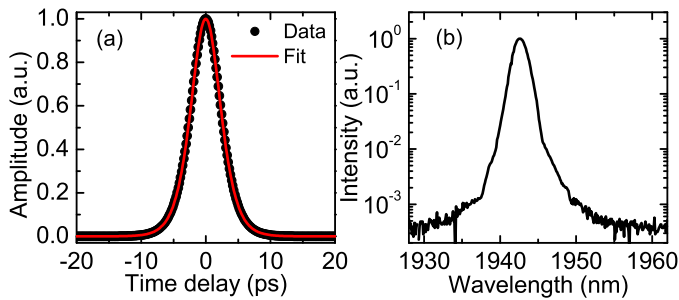}}
\caption{\label{AC+spec} (a) Autocorrelation, (b) optical spectrum.}
\end{figure}

The autocorrelation of the output pulse, and the corresponding optical spectrum are plotted in Fig.\ref{AC+spec}. Fig.\ref{AC+spec}(a) shows that the pulse temporal profile is well represented by a sech$^2$. The FWHM duration (after deconvolution) is 3.6ps. The corresponding FWHM spectral width is 2.1nm, giving a time-bandwidth product$\sim$0.59, indicating low chirp\cite{Agrawal_book_app}. The output power is$\sim$2mW. Although the laser operates with negative cavity dispersion, and the pulses are soliton-like, the typical spectral sideband signature of deviation from average soliton operation is not observed, because the soliton length, given by $z_\mathrm{sol.}=\frac{\pi}{2}\frac{\tau_0^2}{\left|\beta_2\right|}$, is long ($z_\mathrm{sol.}\sim300\mathrm{m}$, based on an estimated $\beta_2=34$~ps~nm$^{-1}$~km$^{-1}$ at 1.94$\mu$m) compared to cavity length (31$~\mathrm{m}\approx\frac{1}{10}z_\mathrm{sol.}$), with $\tau_0$ the pulse duration and $\beta_2$ the group velocity dispersion.

The stability and quality of the generated pulses are evaluated via the radio frequency (RF) spectrum\cite{Linde:86}.
\begin{figure}
\centerline{\includegraphics[width=90mm]{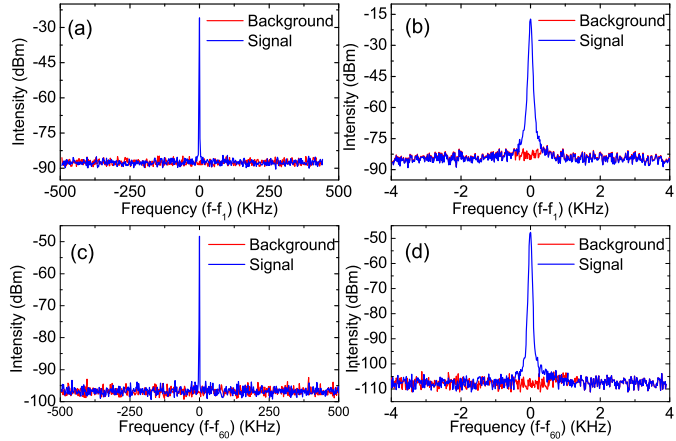}}
\caption{\label{RF} RF spectra. (a) Fundamental, (c) 60$^{th}$  harmonic on a long range span (1MHz), with 300Hz resolution; (b) Fundamental, and (d) 60$^{th}$ harmonic on a short range span (8kHz), with 30Hz resolution.}
\end{figure}

\begin{figure}
\centerline{\includegraphics[width=90mm]{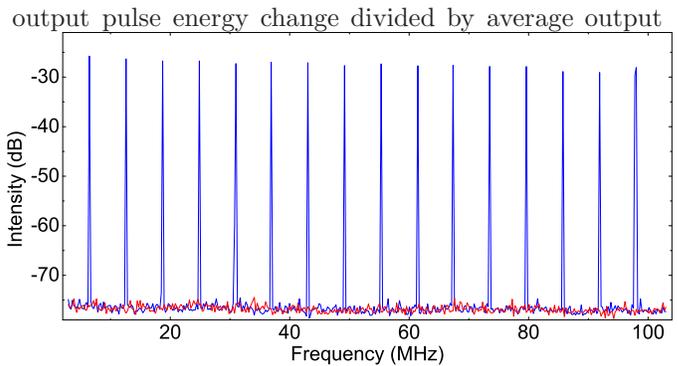}}
\caption{\label{RF2} (Blue) RF spectrum over 100MHz,with 3kHz resolution. (Red) analyzer background}
\end{figure}

Fig.\ref{RF} plots the fundamental and 60$^{th}$ harmonics over long (1MHz) and short (8kHz) frequency spans. The long range spectra indicate that the stability is high, with peak to noise-floor ratio limited by our 300Hz resolution (the noise floor of the analyzer is plotted in red). No sidebands at harmonic cavity frequencies are observed over the 1MHz span, suggesting good pulse-train stability and no Q-switching instabilities. This is confirmed by a spectral sweep over 100MHz, showing the first fifteen harmonics of the fundamental cavity frequency (Fig.\ref{RF2}).

The short range spectra (Fig.\ref{RF}(b,d)), spanning 8kHz with 30Hz resolution, reveal a low level pedestal component$\sim$70dB from the central $f_\mathrm{0}$ spike. Following Ref.\onlinecite{Linde:86}, we estimate the energy fluctuations, defined as output pulse energy change divided by average output energy, as $\Delta E =\left[\frac{\Delta P \Delta f}{\Delta f_\mathrm{Res.}}\right]^{1/2}$, where $\Delta P$ is the power ratio between the central spike at $f_1$ and the peak of the noise band, $\Delta f$ (Hz) is the frequency width of the noise component, and $\Delta f_\mathrm{Res.}$ (Hz) is the resolution bandwidth of the spectrum analyzer. With $\Delta P=1\times10^{-6}$, $\Delta f=730~\mathrm{Hz}$ and $\Delta f_\mathrm{Res.}=30~\mathrm{Hz}$, give low pulse-to-pulse energy fluctuation $\Delta E \approx 5\times10^{-3}$.

Similarly, when the amplitude noise is low, the timing jitter can be evaluated as\cite{Linde:86}:$\frac{\Delta t}{T}=\frac{1}{2\pi n}\left[\frac{\Delta P_n \Delta f}{\Delta f_\mathrm{Res.}}\right]^{1/2}$, where $T$ is the cavity period, $n$ is the harmonic order. The low-frequency timing jitter (Fig.\ref{RF}(d)), evaluated at the 60$^{th}$ harmonic with $\Delta P_{60}=1.6\times10^{-5}$, $\Delta f=393~\mathrm{Hz}$ and $\Delta f_\mathrm{Res.}=30~\mathrm{Hz}$, is estimated as $\Delta t/T=3.9\times10^{-5}$. Given the long cavity period $T=155~\mathrm{ns}$, this indicates a low timing jitter $\Delta t\approx6~\mathrm{ps}$.

Our analysis suggests that, despite a very simple cavity consisting of non-polarization maintaining (PM) fiber, the laser emits high-quality pulses with low amplitude fluctuations ($0.5\%$) and low timing jitter$\sim6\mathrm{ps}$. Although the laser mode-locks without the bandpass filter, the quality of the emitted pulses is compromised, with an increase in the RF spectrum noise.

In summary, we reported stable continuous-wave mode-locking of a Tm-doped fiber laser, using a graphene-based saturable absorber. The laser generates 3.6ps pulses at 6.46MHz, with$\sim$0.4nJ pulse energy, demonstrating the operation of graphene in the mid-IR. This simple all-fiber design supports low noise operation in a small footprint, suitable for packing in a compact single-unit system. This can be realized using all-PM fiber components, which should further improve stability and noise properties\cite{Laegsgaard:08}.

We acknowledge funding from ERC grant NANOPOTS, EPSRC Doctoral Prize Fellowship, EPSRC grant EP/GO30480/1, EU Grants RODIN and GENIUS, a Royal Society Wolfson Research Merit Award, a RAEng Fellowship.

\end{document}